\documentclass[journal]{IEEEtran}
\usepackage{amsmath,amssymb,graphicx}
\usepackage[normalem]{ulem}
\newtheorem{thm}{Theorem}[section]

\def\bbox{{\hfill $\Box$}}

\title{Four-Dimension Cross Constellations with Gray Mapping}

\author{Liangping Ma,
        ~Hae Chung,
        ~Byung K Yi 
\thanks{L. Ma is with InterDigital Communications, Inc., San Diego, CA 92121, USA (e-mail: liangping.ma@interdigital.com).}
\thanks{Hae Chung is with Kumoh National Institute of Technology, Gumi, Korea (e-mail: hchung@kumoh.ac.kr).}
\thanks{Byung K Yi was with InterDigital Communications, Inc., San Diego, CA 92121, USA.}
\thanks{The work was last updated April 2017.}
}

\begin{document}
\maketitle

\begin{abstract}
Recently a four-dimension (4D) cross constellation has been proposed, where a 4D-vector is drawn from two $(3\times 4^m)$-ary QAM constellations, in an effort to reduce the peak-to-average-power ratio (PAPR). We construct a bits-to-signal mapping and prove that it is a Gray mapping. Simulation results show that the proposed modulation scheme is effective in reducing the PAPR while providing better error performance than existing 4D modulation schemes. 
\end{abstract}
%

%
\IEEEpeerreviewmaketitle

\section{Introduction}
\label{sec:intro}
Four-dimension (4D) modulation has been widely used in optical communication~\cite{Opbook}, where a signal is naturally 4-dimension: a quadrature component and an inphase component with two polarizations for each. Polarizations are also leveraged in a proposed 4D modulation for satellite communication~\cite{Taricco93} and more recently for mmWave communication in the context of the fifth generation (5G) communication systems~\cite{Love15}. Even without polarizations, 4D modulation can be realized by doing quadrature amplitude modulation (QAM) over two resources such as two time intervals or two frequencies, for example, as in an OFDM system. One of the motivations for using 4D modulation is an increased energy efficiency over the traditional two-dimension (2D) modulation (i.e., using the quadrature and inphase components only), as packing in a higher-dimension space generally reduces the average signal power needed to achieve a given minimum distance among modulated signals~\cite{Proakis08}.

Energy efficiency is an important design consideration for New Radio (NR)~\cite{NR_RP} -- the radio for 5G. Since the NR waveforms will be based on OFDM as in Long-Term Evolution (LTE)~\cite{36.211} and OFDM may cause high Peak-to-Average-Power Ratio (PAPR), which is detrimental to energy efficiency, it is important to reduce the PAPR of the waveforms, especially for the uplink transmissions where the User Equipments(UEs) have limited power supply. Major PAPR reduction techniques include clipping and peak windowing, but they cause in-band distortion and clipping additionally causes out-of-band emission~\cite{Ch07}\cite{Prasad04}. Thus, new effective and simple methods for reducing the PAPR are worth exploring.

Uniform $4^m$-ary QAM, where $m=1, 2, 3, 4$, has been adopted by LTE~\cite{36.211} and 5G~\cite{NR_RP}. The corner points in such constellations tend to contribute more to the PAPR than other points. Therefore, it may be beneficial to remove certain corner points, resulting in a cross constellation~\cite{Cross62}. For example, removing the 4 corner points of a 16-QAM constellation results in a 12-QAM constellation. In the meantime, it is important to make the bits-to-signal mapping a Gray mapping in order to have good bit error rate (BER) performance. Given the energy efficiency advantage of 4D modulation, it may be beneficial to consider 4D modulation design. There are works on the design of efficient 4D constellations~\cite{Welti74}, but very little on the Gray mapping design for such constellations. In fact, for a cross constellation, the Gray mapping from bits to signals generally is not available~\cite{Cross05}\cite{Milstein99}.
A recent attempt is made in \cite{Chung16}, where a cross constellation in the shape of a 12-QAM is extended twice in frequency or time, resulting in a 4D cross constellation with a modulation efficiency of (2m+1.5) bits/2D which adds to the modulation granularity. 
However, the bits-to-signal mapping in \cite{Chung16} appears overly complicated, without a proof of its validity (i.e., it is indeed a Gray mapping for all cases). In this letter, we take a geometric approach, which is different from \cite{Chung16}, to constructing a mapping, prove that it is a Gray mapping, and evaluate the PAPR and error performance against existing schemes.

Throughout this letter, we adopt the notations in \cite{KimBook}: $x_i^j$ denotes the sequence $x_i, \dots, x_j$, a boldface lower case letter such as $\textbf{x}$ denotes a row vector, and a boldface upper case letter $\textbf{G}$ denotes a matrix. A symbol means a 2D signal, and we use 4D-vector and \emph{2-symbol sequence} interchangeably.

The remainder of the letter is organized as follows. Section \ref{sec:12-ary} presents the case $m=1$ (12-ary QAM) in two complex dimensions, Section \ref{sec:gen} generalizes the result for arbitrary values for $m$, and Section \ref{sec:perf} gives the PAPR performance and error performance. Lastly, Section \ref{sec:con} concludes the paper.

\section{12-ary QAM}
\label{sec:12-ary}
\begin{figure}[!h]
\centering
\includegraphics[width=3in]{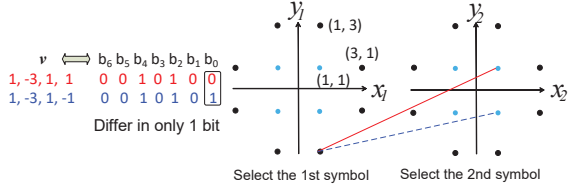}
\caption{A mapping between bit sequences (each of 7 bits) and 4D-vectors.}
\label{fig:12QAM_2d}
\end{figure}

The 12-ary QAM is a $(3\times 4^m)$-ary QAM with $m=1$, and as we will see shortly it serves as a building block for the general case. We start with a 16-QAM constellation and remove the four corner points and get a 12-QAM constellation as shown in Fig.~\ref{fig:12QAM_2d} in the hope of reducing the PAPR. However, a symbol in a single 12-QAM constellation cannot represent an integer number of bits without wasting a significant portion of the symbols. If we take 8 out of 12 symbols to represent 3 bits, we will leave 4 symbols unused, which account for 1/3 of the total symbols. To address this issue, we consider constellation extension, resulting in 144 4D-vectors. We use 128 of these 4D-vectors (or 2-symbol sequences) to represent 7 bits, leaving 16 vectors or a fraction of 1/9 unused. We would like to have a Gray mapping between the 7 bits and the 128 4D-vectors, meaning that if two 4D-vectors are nearest to each other (at distance 2), the corresponding 7-bit sequences differ only in 1 bit. Fig.~\ref{fig:12QAM_2d} shows two 4D-vectors that are closest to each other, indicated by a red line and a blue line, respectively. Since each blue symbol has 4 nearest symbols at distance 2, a 4D-vector consisting of 2 blue symbols has 8 nearest 4D-vectors. However, there are only 7 bits available, and a Gray mapping is infeasible if we select any 4D-vector composed of two blue symbols. By excluding the 16 all-blue 4D-vectors, we fully utilize all legitimate vectors while leaving open the possibility for a Gray mapping.


We define a basic mapping $\textbf{f}(\cdot)$ in Table \ref{tab:1}. It maps 3 bits to two symbols, each of which is taken from the first quadrant of a 12-ary QAM constellation. It can be verified that $\textbf{f}(\cdot)$ is a Gray mapping.

\begin{table*}[t]
\caption{The basic mapping function}
\begin{center}
\begin{tabular}{|c|c|c|c|c|c|c|c|}
\hline
$\textbf{f}(000)$ & $\textbf{f}(001)$ & $\textbf{f}(011)$ & $\textbf{f}(010)$ & $\textbf{f}(110)$ & $\textbf{f}(111)$ & $\textbf{f}(101)$ & $\textbf{f}(100)$\\
\hline
1, 3, 1, 3 & 1, 3, 1, 1 & 1, 3, 3, 1 & 1, 1, 3, 1 & 3, 1, 3, 1 & 3, 1, 1, 1 & 3, 1, 1, 3 & 1, 1, 1, 3 \\
\hline
\end{tabular}
\end{center}
\label{tab:1}
\end{table*}

\begin{thm}
For 12-ary QAM extended in two complex dimensions, the following mapping is a Gray mapping
\begin{equation}
\textbf{v}(b_{0}^{6}) = \textbf{f} (b_{6}, b_{5},b_{4}) \textbf{D}(b_{0}^{3})
\label{eq:12qam}
\end{equation}
where $\textbf{D}(b_{0}^{3}) = \mathrm{diag} ((-1)^{b_{3}}, (-1)^{b_{2}}, (-1)^{b_{1}}, (-1)^{b_{0}})$.
\label{th:12}
\end{thm}
\noindent \textbf{Proof:} Consider any pair of 2-symbol sequences that are separated by the minimum distance of 2. The two 2-symbol sequences differ in only one out of four coordinates with the difference being 2, and consequently either the two first symbols or the two second symbols are different. If the two differing symbols are in the same quadrant of a 12-ary QAM constellation, they differ only in 1 bit since the mapping $\textbf{f}(\cdot)$ is a Gray mapping. Otherwise, the differing coordinates must be +1 and -1 in order to give a distance of 2 between the two 2-symbol sequences, which happens if and only if (i) the coordinate in $\textbf{f} (b_{6}, b_{5},b_{4})$ that causes the difference is equal to 1 for each bit sequence and (ii) the corresponding diagonal entry in $\textbf{D}(b_{0}^{3})$ is $+1$ for one bit sequence and and $-1$ for the other bit sequence. By the definition of $\textbf{D}(b_{0}^{3})$, the two bit sequences differ in only one bit (at Hamming distance 1). \bbox

\section{The General case}
\label{sec:gen}

\begin{figure}[!h]
\centering
\includegraphics[width=2.8in]{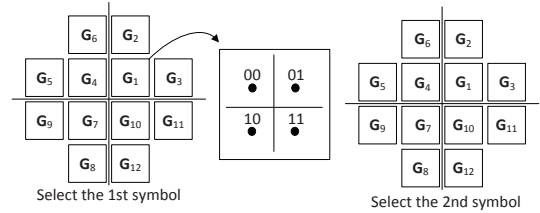}
\caption{A scaled 12-QAM superposed with 4-QAM results in a 48-QAM constellation, which is then duplicated.}
\label{fig:48QAM}
\end{figure}

We now consider $(3\times4^m)$-ary QAM modulation with $m \geq 1$. To have a low PAPR constellation, we derive the $N$-ary QAM constellation by removing a quarter of the points from a $(4^{m+1})$-ary QAM constellation. The resulting constellation can be considered as an enlarged 12-QAM superposed with a $(4^{m-1})$-ary QAM constellation. Using the continuous argument~\cite{Proakis08} we see that the resulting constellation has similar PAPR properties as the 12-QAM. The $(3\times4^m)$-ary QAM constellation is replicated, from which we draw 4D-vectors. We then select a subset of the 4D-vectors and construct a Gray mapping that maps $k$ bits to a 4D-vector, where
\begin{equation}
k=\lfloor \log_2 N^2 \rfloor=3+4m.
\label{eq:k}
 \end{equation}
This modulation carries $k/2=2m+1.5$ bits per 2D, which is not available from a conventional QAM modulation, thus making finer granularity in adaptive modulation and coding (AMC). For example, for $m=2$, we first enlarge a 12-QAM constellation by a factor of 2, and then superpose each point with a $4$-QAM constellation, as shown in Fig.~\ref{fig:48QAM}.

To be a Gray mapping, for any pair of nearest 2-symbol sequences, the bit sequences that they represent differ only in 1 bit. We call each square in Fig.~\ref{fig:48QAM} a \emph{cluster}. The Gray mapping is constructed in two steps: (i) map 7 bits to a 2-symbol sequence drawn from two scaled 12-QAM constellations, where each symbol serve as the center of a cluster, and (ii) map the remaining bits to symbols within the two clusters.

\begin{figure}[!h]
\centering
\includegraphics[width=1.2in]{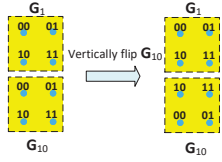}
\caption{Vertically flipping $\textbf{G}_{10}$ eliminates the difference in the partial bit representations for nearest points on the cluster boundary.}
\label{fig_ex}
\end{figure}

\begin{figure}[!h]
\centering
\includegraphics[width=2.7in]{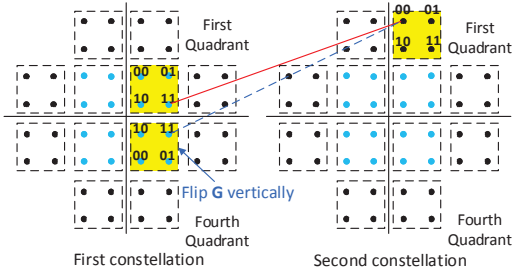}
\caption{Bits-to-signal mapping resulting from the flipping in Fig.~\ref{fig_ex}).}
\label{fig_proof_2}
\end{figure}

To illustrate, we consider a Gray mapping for 48-QAM (i.e., $m=2$) and map $k=11$ bits $\textbf{b}^{10}_0$ to two 48-QAM symbols. In step (i), we map 7 bits $\textbf{b}^{10}_4$ to 2 12-QAM symbols according to (\ref{eq:12qam}) but scaled by a factor of 2. The 2 12-QAM symbols select two clusters from two 48-QAM constellations shown in Fig.~\ref{fig:48QAM}. In step (ii), we map 2 bits $\textbf{b}^{3}_2$ to a point in the first selected cluster and the remaining 2 bits $\textbf{b}^{1}_0$ to a point in the second. Now consider two nearest 4D-vectors. If they are in the same cluster sequence (of two clusters), for example, $(\textbf{G}_{10},~\textbf{G}_2)$, it is sufficient to have the clusters follow a Gray mapping within the respective clusters. On the other hand, if they are in different cluster sequences, the two cluster sequences differ only in one place, at which the clusters are next to each other, for example, $(\textbf{G}_{10},~\textbf{G}_2)$ and $(\textbf{G}_{1},~\textbf{G}_2)$ shown in Fig.~\ref{fig_proof_2} differ in the first cluster and $\textbf{G}_{10}$ and $\textbf{G}_{1}$ are next to each other. If we use the same Gray mapping in the clusters that are different, then on the boundary the nearest two points, one from each cluster, will differ in 1 bit in their partial bit representations $\textbf{b}^{3}_0$, as shown on the left of Fig.~\ref{fig_ex}. If we flip $\textbf{G}_{10}$ vertically, we can eliminate this difference. As a result, the complete bit representations for the nearest 4-D vectors differ only in 1 bit.

Now we make the Gray mapping algorithm more precise and prove its correctness. Let the $k$-bit sequence be $\textbf{b}=b_0^{k-1}$, and the 2-symbol sequence be $\textbf{v}(\textbf{b})=(x_1(\textbf{b}), y_1(\textbf{b}), x_2(\textbf{b}), y_2(\textbf{b}))$, where $x_i(\textbf{b})$ and $y_i(\textbf{b})$ are the coordinates of symbol $i$ for $i=1,2$.
Define
\begin{eqnarray}
\textbf{t}(b_{k-7}^{k-1}) &=& \textbf{f} (b_{k-1}, b_{k-2},b_{k-3}) \textbf{D}(b_{k-7}^{k-4}) \label{eq:t1}\\
&=& (\tilde{x}_1, \tilde{y}_1, \tilde{x}_2, \tilde{y}_2) \label{eq:t}
\end{eqnarray}
where $\textbf{D}(b_{k-7}^{k-4}) = \mathrm{diag} ((-1)^{b_{k-4}}, (-1)^{b_{k-5}}, (-1)^{b_{k-6}}, (-1)^{b_{k-7}})$. The Gray mapping is
\begin{equation}
\textbf{v}(\textbf{b})= 2^{m-1} \textbf{t}(b_{k-7}^{k-1}) + (\textbf{s}(\tilde{x}_1, \tilde{y}_1, b_{(k-7)/2}^{k-8}), \textbf{s}(\tilde{x}_2, \tilde{y}_2, b_0^{(k-9)/2})) \label{eq:map}
\end{equation}
where the first part (to the left of $+$) is from step (i), the remaining part is from step (ii), $\tilde{x}_1, \tilde{y}_1, \tilde{x}_2, \tilde{y}_2$ are defined in (\ref{eq:t}), and
\begin{equation}
\textbf{s}(\tilde{x}_1, \tilde{y}_1, b_{\frac{k-7}{2}}^{k-8}) = \textbf{p}\left(\textbf{U}^{\mathrm{ mod}_2(\frac{|\tilde{y}_1-1|}{2})} \textbf{G} \textbf{U}^{\mathrm{ mod}_2(\frac{|\tilde{x}_1-1|}{2})}, b_{\frac{k-7}{2}}^{k-8} \right)
\label{eq:cluster}
\end{equation}
where $\mathrm{ mod}_2(\cdot)$ stands for the modulo 2 operation, $\textbf{G}$ represents a Gray mapping for an implicit uniform $4^{m-1}$-ary QAM constellation, $\textbf{U}$ is an antidiagonal matrix that performs matrix flipping, and $\textbf{p}$ returns a point from the constellation denoted by $\textbf{U}^{\mathrm{ mod}_2(\frac{|\tilde{y}_1-1|}{2})} \textbf{G} \textbf{U}^{\mathrm{ mod}_2(\frac{|\tilde{x}_1-1|}{2})}$ for bits $b_{\frac{k-7}{2}}^{k-8}$, $\tilde{x}_1$ and $\tilde{y}_1$.

The antidiagonal matrix $\textbf{U}$ has the same dimension as $\textbf{G}$, and it flips a matrix vertically if it left multiplies the matrix and flips a matrix horizontally if it right multiplies the matrix. For example, the $\textbf{U}$ corresponding to a $2 \times 2$ matrix $\textbf{G}$ is
\begin{equation}
\textbf{U} = \left( \begin{array}{cc}  0 & 1 \\  1 & 0 \end{array}  \right)
\end{equation}

\begin{figure}[!h]
\centering
\includegraphics[width=1.3in]{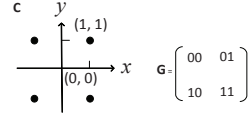}
\caption{An example of constellation $\textbf{C}$ and bit-sequence matrix $\textbf{G}$ for $n=2$.}
\label{fig_C_G}
\end{figure}

The function $\textbf{p}(\textbf{G}, b_{0}^{n-1})$ does Gray mapping and  maps the bit sequence $b_{0}^{n-1}$ to a point of an implicit uniform $2^n$-QAM constellation $\textbf{C}$. Each entry of $\textbf{G}$ is a sequence of $n$ bits and we can think of it as the unsigned binary representation of an integer in the range from 0 to $2^n-1$, as shown for $n=2$ in Fig.~\ref{fig_C_G}. The center of the constellation $\textbf{C}$ is at the origin. 

\begin{figure}[!h]
\centering
\includegraphics[width=1.6in]{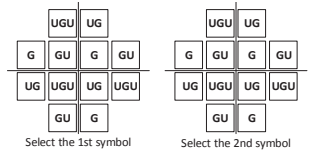}
\caption{A flipping that makes nearest symbols on a cluster boundary have the same partial bit representations.}
\label{fig_flip}
\end{figure}

\begin{thm}
The mapping in (\ref{eq:map}) is a Gray mapping.
\label{th:gen}
\end{thm}
\noindent \textbf{Proof:}
Let the mapping within cluster $\textbf{C}_i$ be $\textbf{G}_i$, which is a Gray mapping, where $i=1, \dots, 12$. For any pair of 2-symbol sequences at the minimum distance 2, they differ in only one symbol and in fact in only one coordinate. The differing symbols both are either in the first constellation or the second. We consider the case where they are in the first constellation (and the argument for the other case is similar), where the two differing symbols (the first symbols of the two 2-symbol sequences) are either in the same cluster or in two clusters that are next to and face each other. If they are in the same cluster, then the $b_{k-7}^{k-1}$ sequences for the two 2-symbol sequences will be the same since $\textbf{t}(\cdot)$ is a one-to-one mapping. Moreover, the $b_{(k-7)/2}^{k-8}$ sequences for the two 2-symbol sequences differ in only 1 bit since the mapping $\textbf{G}_i$ is a Gray mapping. Lastly, the $b_{0}^{(k-9)/2}$ sequences for the two 2-symbol sequences are identical. Thus, the two whole bit sequences for the two 2-symbol sequences differ only in 1 bit. If the two differing symbols are in two clusters that are next to and face each other, e.g., $\textbf{G}_1$ and $\textbf{G}_2$ but not $\textbf{G}_1$ and $\textbf{G}_7$ in Fig.~\ref{fig:48QAM}, then the $b_{k-7}^{k-1}$ sequences for the two 2-symbol sequences will differ in only 1 bit because $\textbf{t}(\cdot)$ is a Gray mapping (see Theorem \ref{th:12} and (\ref{eq:t1})). That requires the $b_{(k-7)/2}^{k-8}$ sequences for the two 2-symbol sequences to be identical. The requirement will not be satisfied if $\textbf{G}_i$'s are the same, i.e., $\textbf{G}_i=\textbf{G}$, where $i=1, \dots, 12$ and $\textbf{G}$ is a Gray mapping. The problem is resolved if we flip the matrix $\textbf{G}$ horizontally, vertically or both. An example is shown in Fig.~\ref{fig_proof_2}, where two nearest 2-symbol sequences are indicated by a red line and a blue dashed line, respectively, and $\textbf{G}_{10}$ (see Fig.~\ref{fig:48QAM}) is initially assigned with $\textbf{G}$ and then flipped vertically. Considering flipping all $\textbf{G}_i$ leads to Fig.~\ref{fig_flip}, where matrix flipping is denoted succinctly by matrix multiplication with an antidiagonal matrix. The resulting matrix is in the form $\textbf{U}^{i_1}\textbf{G}\textbf{U}^{i_2}$, where $i_1=0, 1$ and $i_2=0,1$. Specifically, $\textbf{U}\textbf{G}$ flips $\textbf{G}$ vertically, $\textbf{G}\textbf{U}$ flips $\textbf{G}$ horizontally, $\textbf{U}\textbf{G}\textbf{U}$ flips $\textbf{G}$ both vertically and horizontally. Lastly, $b_{0}^{(k-9)/2}$ for the two 2-symbol sequences are identical. Thus, the two whole bit sequences for the two 2-symbol sequences differ only in 1 bit. This flipping operation does not change the Gray mapping within a cluster, thus still being able to satisfy the Gray mapping requirement for the case where the two differing symbols are in the same cluster. Therefore, the above construction gives a Gray mapping.
Inspecting the matrix exponents in Fig.~\ref{fig_flip} together with (\ref{eq:t}) yields (\ref{eq:cluster}).
\bbox

Note that the proposed Gray mapping, is not balanced, i.e., different bits have different BERs.



\section{Performance evaluation}
For performance comparison, we take the 4D constellation (Class I) of \cite{Welti74}, which has 145 points on a 4D square lattice. For fair comparison, we eliminate the high-power points to get a 128-point constellation. We also compare with the 4D dicyclic group constellation of size 128~\cite{Zet77}, a combination of two PSK constellations.
\label{sec:perf}
\subsection{PAPR reduction}  
\begin{figure}[!h]
\centering
$ \begin{array}{cc}
\includegraphics[width=1.7in]{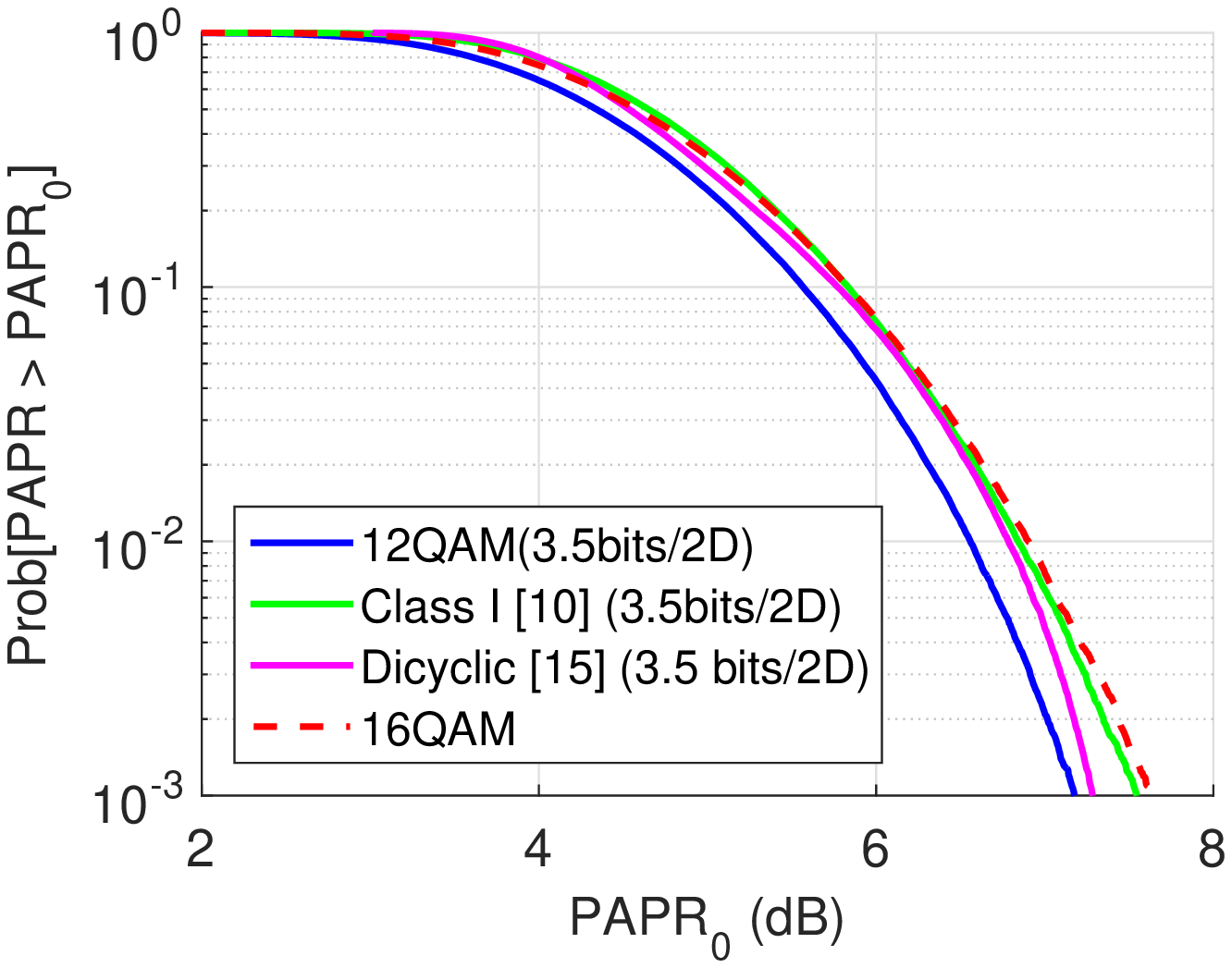} &
\includegraphics[width=1.7in,height=1.2in]{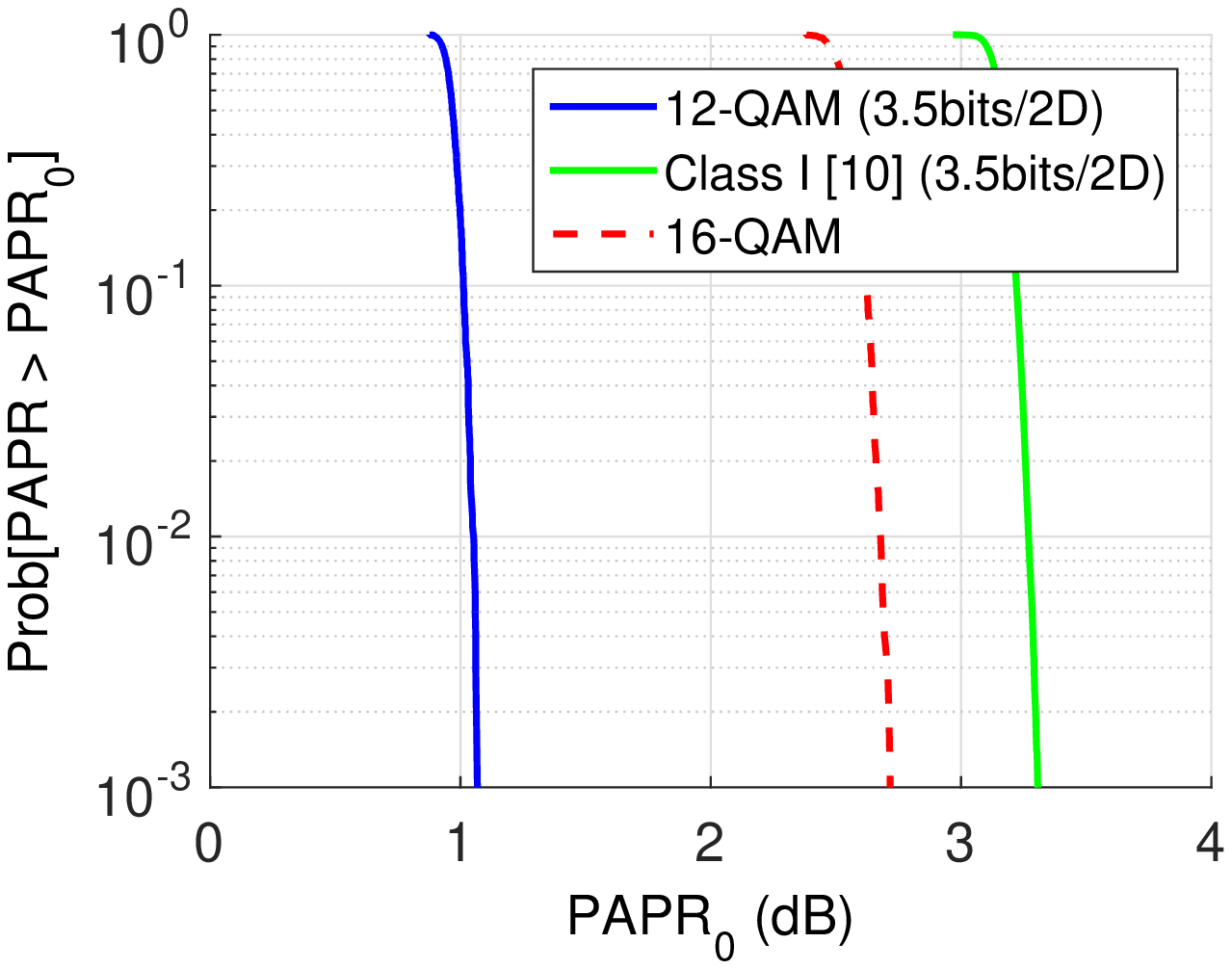} \\
(a) & (b)
\end{array}
$
\caption{PAPR of DFT-spread-OFDM for (a) 12 subcarriers and (b) 2048.}
\label{fig_papr12}
\end{figure}

The PAPR can be evaluated for the constellation alone or for the resulting waveform. For conventional 2D QAM transmissions, the PAPRs for the proposed 12-QAM, Class I~\cite{Welti74}, dicyclic~\cite{Zet77} and 16-QAM are 0.969dB, 3.162dB, 3.000dB and 2.553dB, respectively. For the waveform-based evaluation, we consider DFT-spread OFDM, whose PAPR depends on both the constellation and the numbers of subcarriers (those being used and the total). The simulation results are shown in Fig.~\ref{fig_papr12}, where the total number of subcarriers is 2048. For the case of 12 subcarriers (used in the IFFT stage), 12-QAM outperforms the other schemes by 0.2 to 0.5 dB at probability level $10^{-3}$, as shown in (a). For the case of all 2048 subcarriers, the waveform is purely single-carrier, and the PAPR of the constellation determines the PAPR of the waveform, and the PAPR reduction by the use of 12QAM becomes more significant, as large as 2dB. The PAPR for the dicyclic group modulation~\cite{Zet77} takes a single value of 3dB and does not exhibit a general distribution. The PAPR reductions can be significant in practice.



\subsection{Error performance}
\begin{figure}[!h]
\centering
$ \begin{array}{cc}
\includegraphics[width=1.7in]{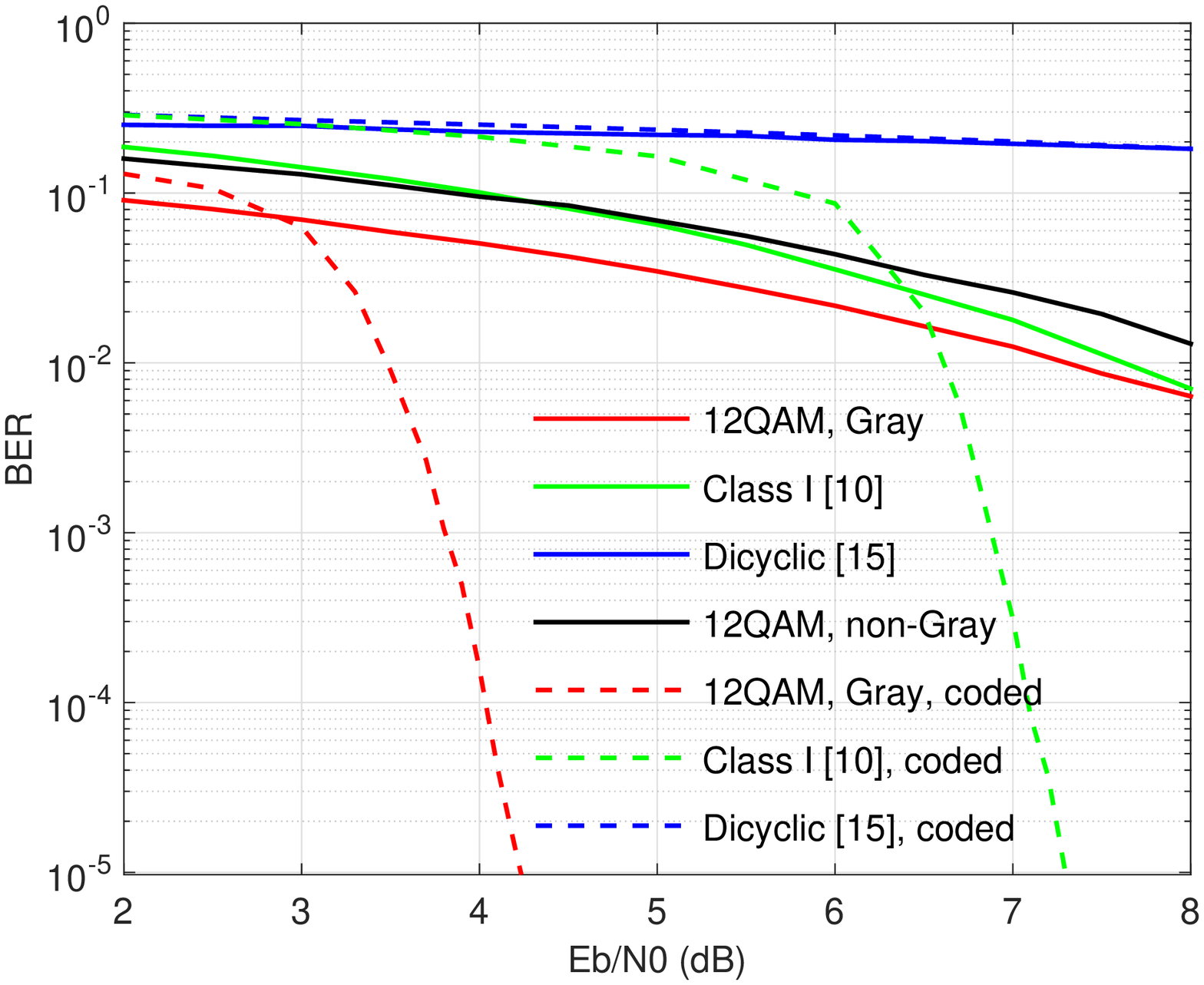} &
\includegraphics[width=1.7in, height=1.27in]{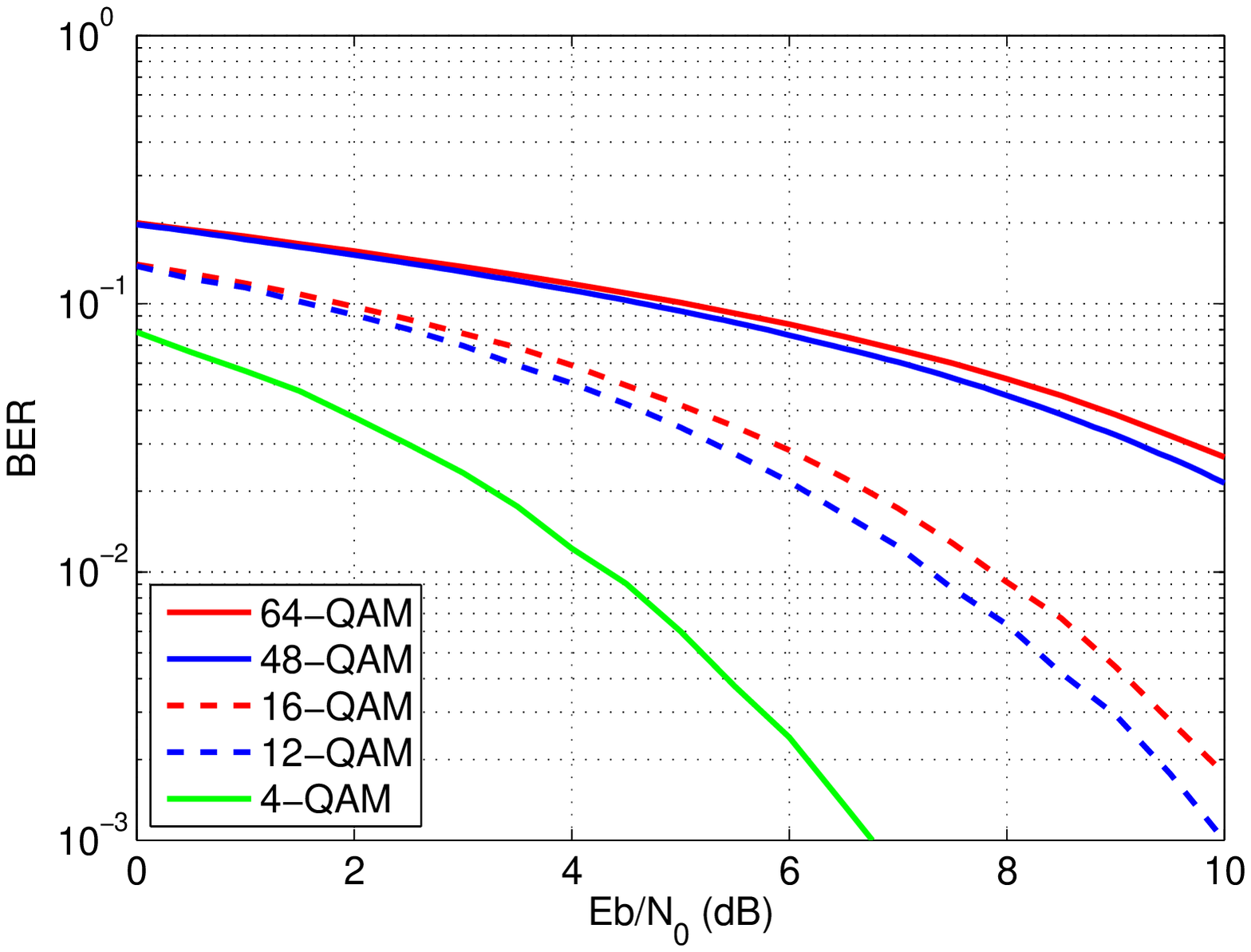} \\
(a) & (b)
\end{array}
$
\caption{(a) BER performance comparison with competing solutions; (b) $(3\times 4^m)$-ary QAM ($m=1,2$) gives a finer granularity in adaptive modulation.}
\label{fig_uncoded}
\end{figure}

The signal packing of the 4D constellation (Class I) of \cite{Welti74} of 128 points is efficient, but does not allow a Gray mapping, because the average number of neighbors at the minimum distance is 13.58 and in fact the maximum is 24 while a 7-bit sequence can have at most 7 Gray sequences. For our proposed 12QAM, the number is 4 or 5. For capacity achieving codes like LDPC, which operates on parallel bit subchannels, our scheme has significant advantages, as shown by the dashed lines from simulations in Fig.~\ref{fig_uncoded} (a), which also shows gains in uncoded performance. The LDPC code has rate 1/2, column weight 3, and blocklength 2394 bits. The channel is AWGN. To show the benefit of the proposed Gray mapping, we evaluate the performance of 12QAM with a non-Gray mapping (the black solid line), which progressively assigns the bit sequences to the 4D points. The dicyclic group modulation~\cite{Zet77} performs poorly because the packing results in a much smaller minimum distance than the other cases. Also, the proposed $(3\times 4^m)$-ary QAM enhances the granularity of modulation, as shown in Fig.~\ref{fig_uncoded} (b).

\section{Conclusion}\label{sec:con}
We presented 4D cross constellations with a Gray mapping, which could significantly reduce the PAPR for OFDM-based waveforms while achieving better error performance than existing 4D modulation schemes.

\bibliographystyle{IEEEbib}

\end{document}